\documentclass[pre,onecolumn,superscriptaddress]{revtex4-1}
\usepackage{graphicx}
\usepackage{amsmath}
\usepackage{amssymb}
\usepackage{booktabs}

\begin{document}

\title{Simplicial closure fragments the explosive cooperation transitions in
higher-order public goods games}
\author{Wenjia Rao}
\email{wjrao@hdu.edu.cn}
\affiliation{School of Sciences, Hangzhou Dianzi University, Hangzhou 310018, China}

\author{Jianneng Huang}
\affiliation{School of Sciences, Hangzhou Dianzi University, Hangzhou 310018, China}

\begin{abstract}
Hypergraph (HG) and simplicial complex (SC) are two common representations
of higher-order networks, and are often expected to differ mainly
quantitatively when they encode comparable group interactions. Here we show
that this expectation fails in evolutionary cooperation dynamics. Using a
controlled higher-order public goods game (PGG) with minimal ad hoc
parameters, we compare cooperation transitions on randomized HG and SC
constructed from the same triangular backbone. We find that the impact
of simplicial closure is selective: when the cooperation transition is continuous-like on HG, imposing simplicial closure mainly broadens the transition without changing its qualitative nature; however, when the transition is explosive and first-order-like on HG, simplicial closure will fragment the compact low/high bistability into a broad ensemble of metastable final states. Detailed analysis reveals that this
selective effect arises from the dual role of simplicial closure in cooperative-nucleus
dynamics: it promotes the survival of local cooperative nuclei while suppressing their conversion into
system-wide cascades. Further experiments confirm that this fragmentation is
not tied to a specific payoff form, but is a generic feature of explosive
cooperation transitions in higher-order PGGs.
\end{abstract}

\maketitle

\section{Introduction}

Higher-order networks provide a natural framework for describing collective
interactions that cannot be decomposed into independent pairwise links. Two
widely used representations are hypergraphs (HG) and simplicial complexes
(SC). In an HG, a higher-order interaction can connect an arbitrary set of
nodes, whereas an SC imposes simplicial closure: every higher-order simplex
necessarily contains all of its lower-order faces\cite%
{Bianconi2021,Bick2023,Torres2021,Battiston2020,Benson2018,battiston2021,boccaletti2023,majhi2022}%
. Since both representations can encode group interactions, they are often
treated as qualitatively similar when the number and size of higher-order
interactions are matched. Recent studies, however, have shown that this
equivalence can break down in certain complex systems. For instance, in
higher-order contagion, simplicial closure can change spreading thresholds
and outbreak patterns by constraining how group exposure is organized locally%
\cite{FerrazdeArruda2024,Iacopini2019,Burgio2024,maia2026}. Another example
is the synchronization dynamics, where SC and HG can lead to different
collective behavior because closure correlates higher-order interactions
with the underlying pairwise structure and changes the effective
distribution of higher-order feedback\cite%
{Lucas2020,Gambuzza2021,Millan2020,zhang2023}. These results suggest that
the distinction between HGs and SCs is not merely a matter of
representation. Nevertheless, how such simplicial closure affects
evolutionary game dynamics remains incompletely understood.

A natural setting to address this question is the public goods game
(PGG), a paradigmatic model of evolutionary cooperation \cite%
{nowak2006,hauert2006}. In the original PGG, cooperators contribute to a
common pool at a personal cost, whereas defectors make no contribution but
still share the collective benefit. This creates a typical social dilemma,
in which the individually optimal choice conflicts with the collectively
beneficial outcome. Many mechanisms have since been introduced into PGGs
to promote cooperation,
including network reciprocity, punishment, reputation, and heterogeneity.
These extensions have made the PGG a minimal yet versatile model for
studying the rich collective behavior in complex systems \cite%
{rand2009,santos2008,perc2017,lieberman2005,ohtsuki2006,santos2005,szabo2007,allen2017}%
. In particular, the incorporation of higher-order interactions into PGGs
has attracted increasing attention in recent years \cite%
{civilini2021,alvarez2021,guo2021,civilini2024,sadekar2025,llabres2026,Tang2026,ma2025,hwang2025,zhang2025,wang2026csf}%
. We therefore use the higher-order PGG as a dynamical testbed to examine
how simplicial closure affects evolutionary dynamics.

In this work, we study a controlled PGG with both pairwise and three-player
interactions. Nonlinear collective feedback is introduced only through the
three-player payoff, which contains a single tunable exponent that controls
the scale effect of group returns. By varying this exponent, the same PGG
framework can generate distinct cooperation-transition regimes on randomized
HG, ranging from gradual continuous-like transitions to explosive
first-order-like transitions. For the network representation, we construct
HG and SC from the same parameter-free triangular growth rule, keeping the
pairwise backbone, the number of three-player groups, and the interaction
density matched. The two representations therefore differ only in whether
the three-player interactions obey simplicial closure. This controlled
setting allows us to examine, through a variety of diagnostics, how
simplicial closure affects cooperation dynamics in different transition
regimes.

The central finding of this work is that the effect of simplicial closure on
cooperation dynamics is selective. When the corresponding randomized HG
exhibits a continuous-like cooperation transition, imposing simplicial
closure mainly produces quantitative broadening of the transition, without
changing its underlying transition type. In sharp contrast, when the
randomized HG exhibits an explosive first-order-like transition, simplicial
closure fragments the compact low/high bistability of the HG into a broad
ensemble of metastable final states in the SC. Through detailed microscopic
analysis, we show that this selective effect originates from the dual role
of simplicial closure in cooperative-nucleus dynamics: it enhances the local
persistence of cooperative nuclei, but constrains their propagation into
system-wide cooperative cascades. We further show that this fragmentation
mechanism persists for different three-player payoff forms, indicating that
it is not tied to a specific payoff structure but reflects a generic effect
of simplicial closure on explosive cooperation transitions.

This paper is organized as follows. Section~II introduces the higher-order
PGG and the construction of the HG and SC representations. Section~III
presents the phenomenology of the selective influence of simplicial closure
on cooperation transitions. Section~IV analyzes its microscopic mechanism.
Section~V examines the robustness of this mechanism under partial breaking
of simplicial closure. Section~VI studies the triple-product-form payoff as
an additional test of the proposed mechanism. Section~VII concludes the
paper and discusses broader implications.

\section{Model and network representations}

In this work, we study a higher-order public goods game (PGG) that contains
both pairwise and three-player interactions. Each player $i$ carries a
binary strategy $s_{i}\in \{0,1\}$, with $s_{i}=1$ corresponding to
cooperation and $s_{i}=0$ corresponding to defection. In every local game, a
cooperator pays a unit cost, while a defector makes no contribution. For a
pairwise interaction between players $i$ and $j$, the net payoff
contribution assigned to player $i$ is
\begin{equation}
u_{i}^{(2)}(i,j)=\frac{r}{2}(s_{i}+s_{j})-s_{i},  \label{eq:pair_payoff}
\end{equation}%
where $r$ denotes the multiplication factor. For a three-player interaction
on a 3-hyperedge $(i,j,k)$, we write $S_{ijk}=s_{i}+s_{j}+s_{k}$ for the
number of cooperators in the group. The payoff contribution received by
player $i$ from this three-player game is formulated as
\begin{equation}
u_{i}^{(3)}(i,j,k)=3\alpha r\left( \frac{S_{ijk}}{3}\right) ^{\gamma }-s_{i}.
\label{eq:three_payoff}
\end{equation}%
Here, $\alpha $ sets the relative weight of the three-player interaction,
whereas $\gamma $ determines the curvature of the collective return. The normalization by $S_{ijk}/3$ keeps the maximal collective benefit of a fully cooperative triad fixed when $\gamma$ is varied.
Consequently, varying $\gamma $ primarily changes the marginal-return
structure of the three-player payoff, rather than changing the maximal group
benefit itself.

The payoff of player $i$ is computed by adding all pairwise and three-player
contributions involving that player and then normalizing by the total number
of local games in which the player participates:
\begin{equation}
U_i= \frac{ \sum_{j\in \mathcal{N}_i}u_i^{(2)}(i,j) + \sum_{(j,k)\in
\mathcal{T}_i}u_i^{(3)}(i,j,k) } {d_i^{(2)}+d_i^{(3)}} .
\label{eq:total_payoff}
\end{equation}
Here, $\mathcal{N}_i$ is the set of pairwise neighbors of player $i$, and $%
\mathcal{T}_i$ is the set of unordered pairs $(j,k)$ for which $(i,j,k)$
forms a three-player interaction. We denote $d_i^{(2)}=|\mathcal{N}_i|$ and $%
d_i^{(3)}=|\mathcal{T}_i|$ as the numbers of pairwise and three-player games
involving player $i$, respectively. This normalization avoids a trivial
payoff advantage for players that participate in more local games.
The strategy updating is implemented through the Fermi imitation rule.
During one elementary update, a focal player $i$ is first chosen at random,
and a model player $j$ is then selected from the union of $i$'s pairwise
neighbors and three-body partners. The focal player adopts the strategy of
the model player with probability
\begin{equation}
W(s_i\leftarrow s_j)= \frac{1}{1+\exp[(U_i-U_j)/\kappa]} ,  \label{eq:fermi}
\end{equation}
where $\kappa$ controls the noise strength in imitation, unless otherwise stated, we use $\kappa=1.0
$.

This PGG is able to generate two kinds of cooperation transitions on HG:
a continuous-like transition for $\gamma < 1$ and an explosive first-order-like
transition for $\gamma>1$. The underlying distinction comes from the marginal return of
adding cooperators to a three-player group\cite{llabres2026,Tang2026}. For the convex regime $\gamma>1$, the marginal
benefit increases with the number of cooperators, so fully cooperative triads
become disproportionately more rewarding than mixed cooperative-defective
triads. This creates a positive feedback: once sufficiently many fully
cooperative triads are formed, they can self-reinforce and trigger a
cascade-like expansion of cooperation, leading to bistability and an
explosive transition. For concave regime $\gamma<1$, by contrast, the marginal benefit
decreases as the group becomes more cooperative. Mixed local configurations
can therefore persist, and cooperation grows more gradually without a
self-sustaining cascade.

We next construct the HG and SC representations from the same triangular
backbone. The backbone is generated by a triangle-growth procedure: starting
from a fully connected core of $M_0=3$ nodes, each new node is connected to the
two endpoints of a randomly selected existing edge, thereby creating one new
triangle. This parameter-free procedure is repeated until the system reaches
$N$ nodes, yielding a sparse pairwise network with average degree
$\langle k\rangle\to 4$ and $T=N-2$ triangles. The SC representation uses these
closed triangles as three-player groups, while the randomized HG rewires these
groups at random. In this way, the HG and SC representations have identical pairwise and three-player interaction densities, so that any difference in the dynamics can
be attributed to simplicial closure rather than to the number of interactions.

In the following sections, we take $\gamma=0.5$ and $\gamma=2$ as
two representative cases, which respectively generate
continuous-like and first-order-like transitions on HG.
This allows us to compare whether simplicial closure acts differently on
different transition regimes. For self-containedness,
Sec.~III also provides the minimal phenomenological diagnostics of these two
regimes in the present HG--SC comparison.

\section{Regime-dependent effects of simplicial closure}

\begin{figure}[htbp]
\centering
\includegraphics[width=0.8\textwidth]{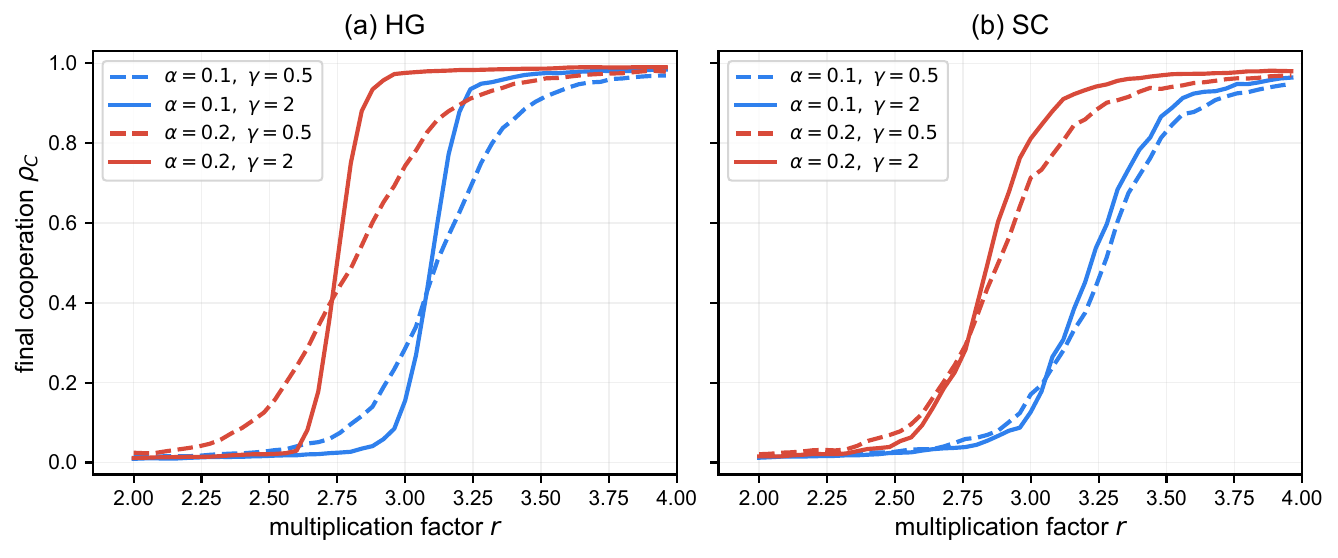}
\caption{Final cooperation ratio $\rho_C$ as a function of the multiplication factor $r$ in (a) randomized hypergraph (HG) and (b) simplicial
    complex (SC).  Results are shown for two payoff exponents,
    $\gamma=0.5$ and $\gamma=2$, and two strengths of the three-body
    interaction, $\alpha=0.1$ and $\alpha=0.2$.  In the HG, the concave case
    $\gamma=0.5$ exhibits a gradual continuous-like transition, whereas the
    convex case $\gamma=2$ leads to a much sharper, discontinuous-like
    transition.  In the SC, the concave regime remains qualitatively similar,
    but the convex regime becomes much smoother. For visual clarity, a moving average
    with a window of three adjacent $r$ values is applied to better display the shapes of the transition curves.} \label{fig:fig1}
\end{figure}

We first compare the mean final cooperation level $\rho_C$ in randomized
HG and SC. Fig.~\ref{fig:fig1} shows $%
\rho_C$ as a function of the multiplication factor $r$ for two
representative payoff regimes, $\gamma=0.5$ and $\gamma=2$, where two values
of the three-body strength $\alpha=0.1,0.2$ are displayed. Each curve is averaged over 100 independent realizations, and each realization is evolved for $1.5\times 10^5$ elementary update steps to ensure convergence.

As expected, increasing $\alpha$ generally promotes the cooperation by lowering
the threshold values of $r$ in all cases. The more important feature, however, is the shape of
the transition curve. In the randomized HG, shown in Fig.~\ref{fig:fig1}(a),
the $\gamma=0.5$ case exhibits a gradual increase of cooperation,
whereas the $\gamma=2$ case shows a much sharper, switch-like
transition. This provides a first indication that the concave and convex
payoffs correspond to continuous-like and first-order-like transition
regimes, respectively.
In the SC, shown in Fig.~\ref{fig:fig1}(b), simplicial closure affects these
two regimes in different ways. For $\gamma=0.5$, the qualitative shape of
the transition is largely preserved, with only minor shifts and broadening.
By contrast, for $\gamma=2$, the transition region is strongly broadened and
the transition curve becomes much smoother. This indicates that simplicial
closure does not merely shift the transition point, but modifies the
organization of the explosive transition itself.

\begin{figure}[htbp]
\centering
\includegraphics[width=0.8\textwidth]{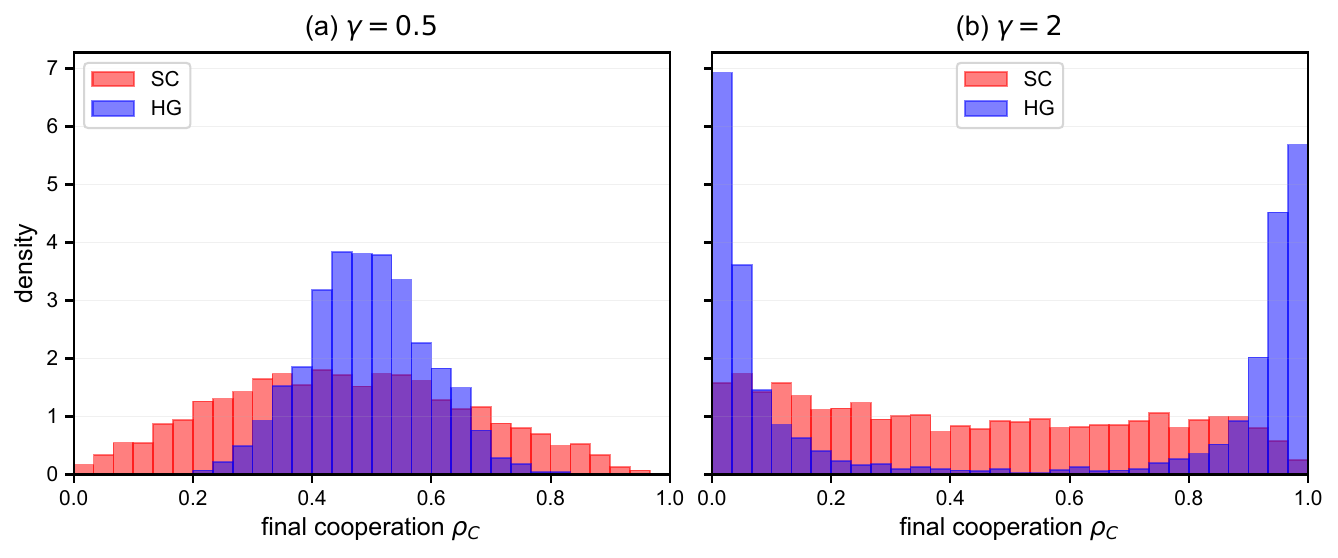}
\caption{Distributions of final cooperation ratio $\rho_C$ near the transition region for (a) $\gamma=0.5$ and (b) $\gamma=2$. For $\gamma=0.5$, both
    representations produce single-peaked distributions centered at
    intermediate cooperation levels, although the SC distribution is broader.
    For $\gamma=2$, the HG exhibits a strongly bimodal distribution,
    indicating global bistability. In contrast, the SC produces a broad distribution over
    intermediate cooperation levels, showing that simplicial closure
    fragments the global bistability of the HG into a broad ensemble of
    metastable final states.} \label{fig:fig2}
\end{figure}

To support the above observation, we further examine the distribution of
final cooperation levels near the transition region. We focus on $\alpha=0.2$
and for each $\gamma$ case, the multiplication factor $r$ is chosen to be
closest to the transition point, namely $r_c=2.78/2.86$ for $\gamma=0.5$ on HG/SC and
 $r_c=2.75/2.82$ for $\gamma=2$ on HG/SC, respectively. The distributions are obtained from 3000
independent realizations with random initial conditions. The results are
displayed in Fig.~\ref{fig:fig2}, where we plot the $\gamma=0.5$ and $\gamma=2$ cases
separately, so that the difference between HG and SC can be compared more
directly.

As we can see, in the $\gamma=0.5$ case, both HG and SC produce single-peaked
distributions centered at intermediate values of $\rho_C$. The distribution
in SC is broader, but the shape stays the same. This confirms that, in the
concave regime, simplicial closure mainly broadens the range of final
outcomes without changing the continuous-like character of the transition.
In contrast, the situation becomes qualitatively different for $\gamma=2$.
In the HG, the distribution is strongly bimodal, with most realizations
ending either near the fully defective state or near the fully cooperative
state, indicating a compact two-basin structure in first-order-like transition. In the SC,
however, the bimodal distribution is replaced by a broad distribution over a
wide range of $\rho_C$. This implies that simplicial closure does not simply
suppress the explosive transition; instead, it fragments the global low/high
bistability of the HG into a broad set of intermediate metastable states. At
the very least, the transition in the SC no longer exhibits the compact
first-order-like structure observed in the randomized HG.

\begin{figure}[htbp]
\centering
\includegraphics[width=0.8\textwidth]{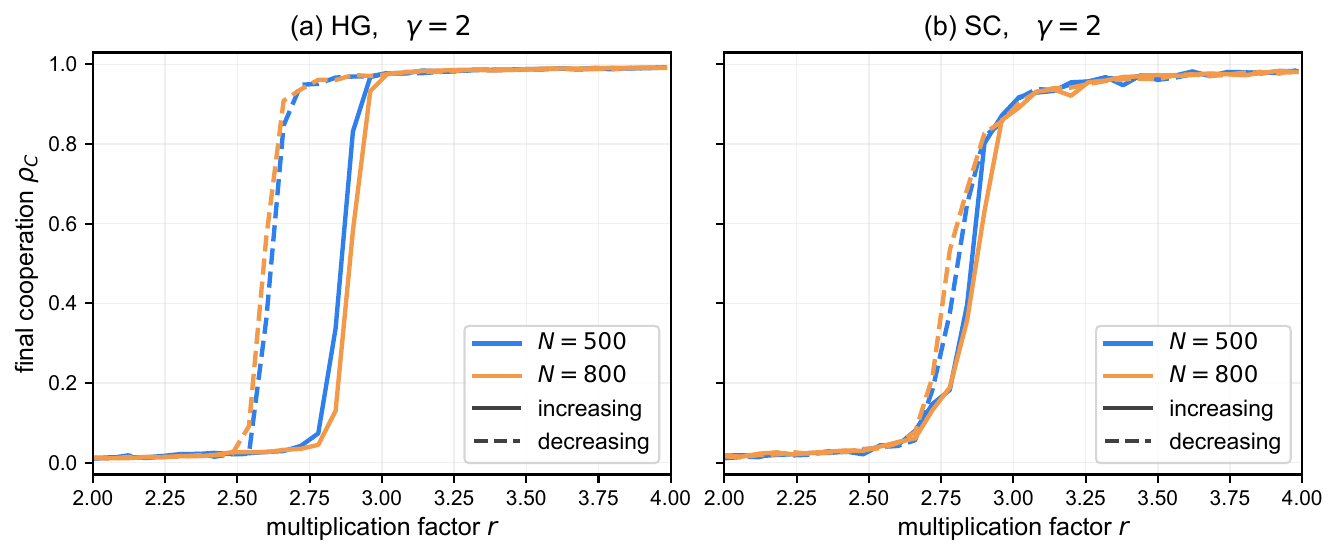}
\caption{Hysteresis scans for the convex payoff regime $\gamma=2$ in
    (a) HG and (b) SC.  Solid lines denote scans with increasing
    $r$, and dashed lines denote scans with decreasing $r$. The HG displays a pronounced
    hysteresis loop, consistent with a first-order-like transition and global
    bistability. In the SC, the hysteresis loop is strongly reduced but does
    not disappear, indicating that simplicial closure does not simply turn the
    explosive transition into an ordinary continuous transition. Instead, it
    produces a smeared discontinuous-like regime with residual path dependence.}
\label{fig:fig3}
\end{figure}

We next perform hysteresis scans to test the history dependence of the
transition, which is a standard dynamical diagnostic of first-order-like
behavior. Specifically, we focus on $\alpha=0.2$ and carry out forward scans
by increasing the multiplication factor $r$ from a low-cooperation state,
and backward scans by decreasing $r$ from a high-cooperation state, both
with an interval $\delta r=0.02$. In each scan direction, the final
configuration obtained at one value of $r$ is used as the initial condition
for the next simulation. For the concave case $\gamma=0.5$, no visible
hysteresis is observed, supporting a continuous-like transition. Therefore,
in Fig.~\ref{fig:fig3} we only show the convex case $\gamma=2$, where each
curve is averaged over 100 independent realizations.

As we can see, in the randomized HG case, the forward and backward scans
follow clearly separated branches over a finite interval of $r$, forming a
pronounced hysteresis loop. Moreover, the loop does not shrink when the
system size is increased from $N=500$ to $N=800$. Together with the bimodal
final-state distribution in Fig.~\ref{fig:fig2}(b), this confirms that the
HG exhibits an explosive first-order-like transition in the convex case $%
\gamma=2$. In contrast, the hysteresis loop is much narrower in the SC,
consistent with the broad rather than bimodal final-state distribution in
Fig.~\ref{fig:fig2}(b). Nevertheless, the two scan directions still do not
collapse onto a single curve, and the residual path dependence remains
visible at the larger system size. This indicates that simplicial closure
does not simply convert the explosive HG transition into an ordinary
continuous transition. Instead, it weakens the global two-basin bistability
while retaining memory effects, leading to a broadened transition region
with many intermediate metastable outcomes.

Taken together, the above results establish the main phenomenological
message of this work: the difference between HG and SC is regime dependent.
For a continuous-like transition, simplicial closure mainly produces
quantitative changes. For an explosive first-order-like transition in the
randomized HG, simplicial closure qualitatively restructures the transition
region: the compact low/high bistability of the HG is replaced by broad
final-state distributions with residual hysteresis, revealing a regime that
is neither an ordinary continuous transition nor a conventional first-order
transition. To clarify the nature of this unconventional transition, we turn
to microscopic diagnostics in the next section.

\section{Fragmentation of Bistability: Metastable States and Cooperative
Nuclei}

The phenomenological results above show that the cooperation transition in
the convex regime on SC is something besides a conventional first-order or
continuous transition. Indeed, the broad distribution of final cooperation
levels suggests that the system may not be attracted to only one or two
macroscopic final states. Instead, it may contain a distributed set of
metastable final states. To test this possibility, we compute the
distribution of the ``final-state overlap''. For two independent
realizations $a$ and $b$, the final-state overlap is defined as
\begin{equation}
Q_{ab}=\frac{1}{N}\sum_{i=1}^{N}\delta \left(s_i^{(a)},s_i^{(b)}\right),
\end{equation}
where $s_i^{(a)}\in\{0,1\}$ is the final strategy of node $i$ in realization
$a$, and $\delta(x,y)$ is the Kronecker delta. Thus, $Q_{ab}=1$ means that
the two final configurations are identical, whereas $Q_{ab}\simeq 1/2$
indicates that they are largely uncorrelated at the node level. Then, its
distribution $P(Q)$ is obtained by randomly sampling $10^4$ pairs from the
3000 final configurations generated from independent random initial
conditions.

\begin{figure}[htbp]
\centering
\includegraphics[width=0.8\textwidth]{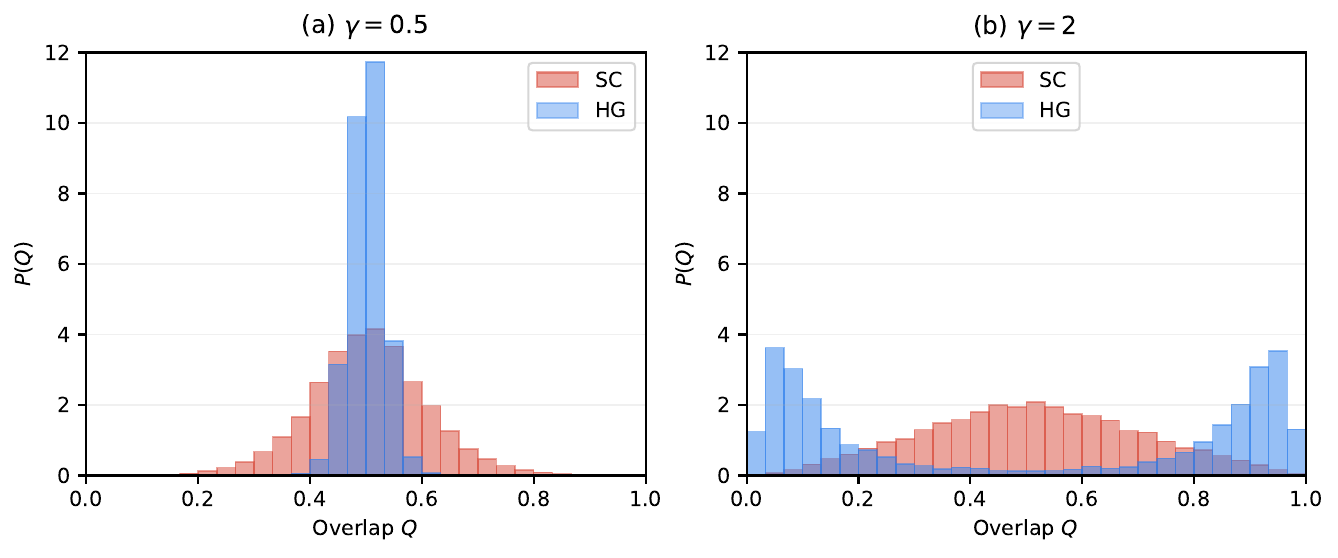}
\caption{Distributions of the final-state configuration overlap $Q$ for
(a) $\gamma=0.5$ and (b) $\gamma=2$. The overlap is computed from randomly
sampled pairs of independently evolved final strategy configurations. For
$\gamma=0.5$, both HG and SC exhibit single-peaked overlap distributions
centered at intermediate values; the SC distribution is broader but remains
regular and bell-shaped. For $\gamma=2$, the HG shows strong weight near both
low and high overlap values, consistent with a compact two-basin structure.
In contrast, the SC distribution becomes broad and flatter over intermediate
overlaps, indicating a distributed set of distinct metastable final
configurations induced by simplicial closure.} \label{fig:fig4}
\end{figure}

It is important to distinguish $P(Q)$ from the distribution of final
cooperation levels $P(\rho_C)$. The latter only describes how many
cooperators remain in the final state, but does not tell us where these
cooperators are located. Two realizations may have the same $\rho_C$ while
involving different sets of cooperative nodes. In contrast, $P(Q)$ measures
the configuration-level similarity between final states. Similar overlap
observables are widely used in glassy and complex systems to probe whether
different realizations converge to the same metastable configuration or to
distinct regions of the attractor landscape. The numerical results for $P(Q)$
are shown in Fig.~\ref{fig:fig4}.

For the concave regime $\gamma=0.5$ in Fig.~\ref{fig:fig4}(a), both HG and
SC produce overlap distributions centered around intermediate values. The HG
distribution is sharply localized near $Q\simeq 0.5$, while the SC
distribution is broader but still has a regular single-peaked, bell-shaped
profile. This is consistent with the final-state distributions in Fig.~\ref%
{fig:fig2}(a): simplicial closure increases the variability of final
configurations, but it does not create a clear separation into distinct
macroscopic basins. The transition therefore remains continuous-like at the
configuration level.

The convex regime $\gamma=2$ in Fig.~\ref{fig:fig4}(b) shows a much stronger
contrast. In the HG, $P(Q)$ has large weight near both low and high overlap
values. This is the overlap signature of a compact two-basin structure:
pairs of realizations falling into the same basin have large overlap,
whereas pairs falling into different basins have low overlap. In the SC,
however, this two-sided structure is replaced by a broad distribution
centered at intermediate overlaps. Compared with the concave case, the SC
distribution is not only broader but also much flatter and less
Gaussian-like, indicating that the broad distribution of $\rho_C$ is not a
simple macroscopic fluctuation around a typical final state. Instead,
independent realizations reach many distinct microscopic final
configurations, showing that simplicial closure fragments the global
bistability of the HG into a distributed set of metastable final states.

The overlap analysis characterizes the structure of final configurations,
but it does not show how these distinct configurations are generated
dynamically. We therefore track the time evolution of the largest
cooperative component in the explosive regime. Let $S_{\max}^{CC}(t)$ denote
the size of the largest connected component composed of cooperative nodes at
time $t$, measured on the pairwise backbone. The normalized quantity $%
S_{\max}^{CC}(t)/N$ indicates whether cooperation remains confined to local
domains or grows into a system-spanning cooperative cluster. Figure~\ref%
{fig:fig5} shows a time-resolved probability-density map of $%
S_{\max}^{CC}(t)/N$ for $\gamma=2$, where the color represents the
probability of observing a trajectory at a given component size and time.

\begin{figure}[htbp]
\centering
\includegraphics[width=0.8\textwidth]{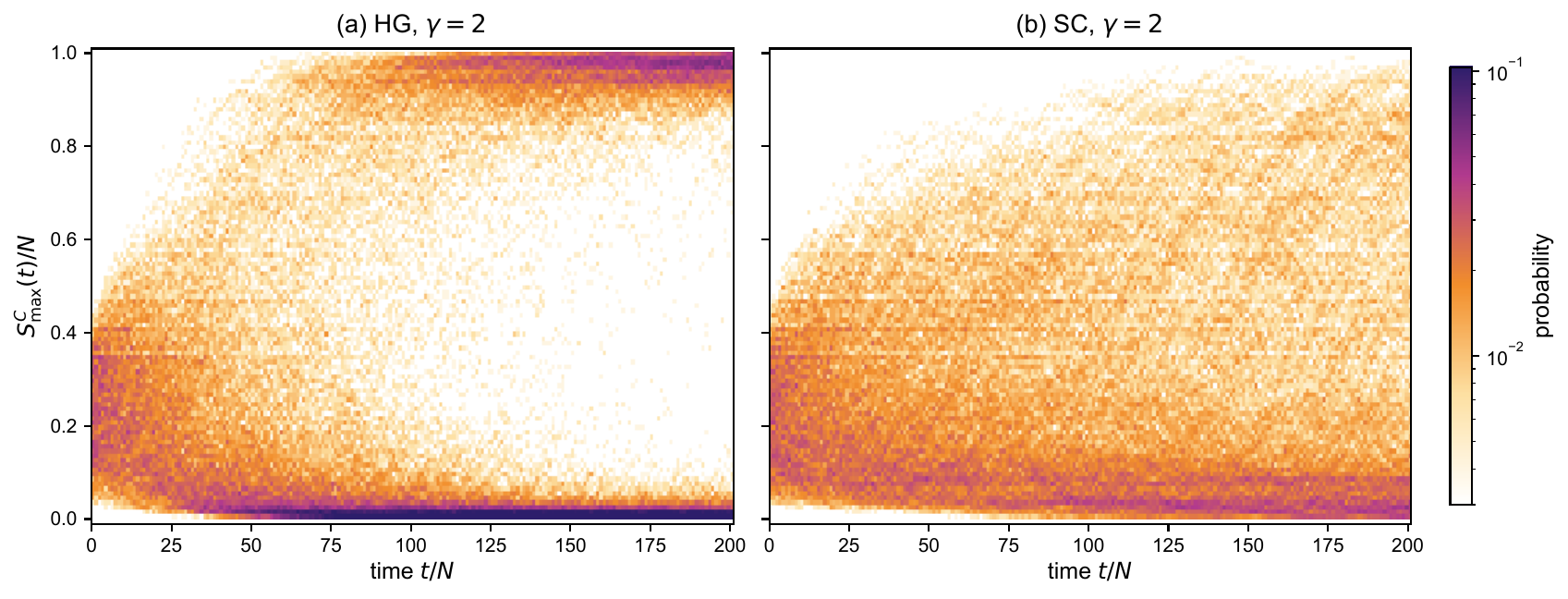}
\caption{Time-resolved probability-density map of the normalized largest cooperative
component $S_{\max}^{CC}(t)/N$ in the convex regime $\gamma=2$. Panel (a) shows
the HG, and panel (b) shows the SC. In the HG, trajectories rapidly separate
into two dominant branches: collapse of cooperative clusters or growth into a
system-spanning cooperative component. In the SC, substantial probability
remains over a wide range of intermediate component sizes, showing that
cooperative nuclei can persist locally for long times without necessarily developing into a
global cascade.} \label{fig:fig5}
\end{figure}

As we can see, in the HG in Fig.~\ref{fig:fig5}(a), the trajectory density
of $S_{\max}^{CC}(t)/N$ rapidly separates into two dominant branches. In
many realizations, the largest cooperative component shrinks to a very small
size, whereas in successful realizations it grows rapidly and approaches the
system size. The intermediate region has relatively low probability,
indicating that cooperative growth in the HG is organized as an all-or-none
process: once a cooperative nucleus crosses a sufficient scale, it tends to
trigger a system-wide cascade.
For the SC in Fig.~\ref{fig:fig5}(b), the evolution of the largest cooperative
component is markedly different. Although small cooperative components can
still disappear and large cooperative domains can still form, the trajectory
density does not collapse into two sharply separated branches. Instead,
substantial probability remains over a wide range of intermediate component
sizes for long times. This shows that simplicial closure broadens the
dynamical routes of cooperative growth: cooperative nuclei can persist and
partially expand without being immediately amplified into a single
system-spanning cascade.

To understand the observations above, we need to examine what simplicial
closure changes at the local structural level. In a randomized HG,
three-body interactions are placed independently of whether the three nodes
form closed triangles in the pairwise backbone. The resulting higher-order
feedback is therefore effectively delocalized. In an SC, by contrast,
three-body interactions are precisely tied to closed triangles, so that
higher-order feedback is organized within overlapping triangular
neighborhoods rather than randomly mixed across the network. This creates
spatially heterogeneous local support: nodes embedded in many closed
three-body groups can provide stronger local reinforcement for cooperative
seeds. At the same time, because the same feedback is constrained to local
triangular neighborhoods, the propagation of such seeds is also expected to
be more spatially confined. This suggests a dual role of simplicial closure:
(1) it may stabilize small cooperative nuclei locally; (2) it reduces the
probability that such locally persistent nuclei are converted into
system-wide cascades.

To test the above structural interpretation, we design a seed experiment
that probes two aspects of cooperative-nucleus dynamics. The first is local
survival: whether small clusters of cooperative seeds can avoid
extinction and remain in the final state. The second is global cascade
conversion: whether seeded cooperation can be amplified into a
system-wide cooperative state.

\begin{figure}[htbp]
\centering
\includegraphics[width=0.8\textwidth]{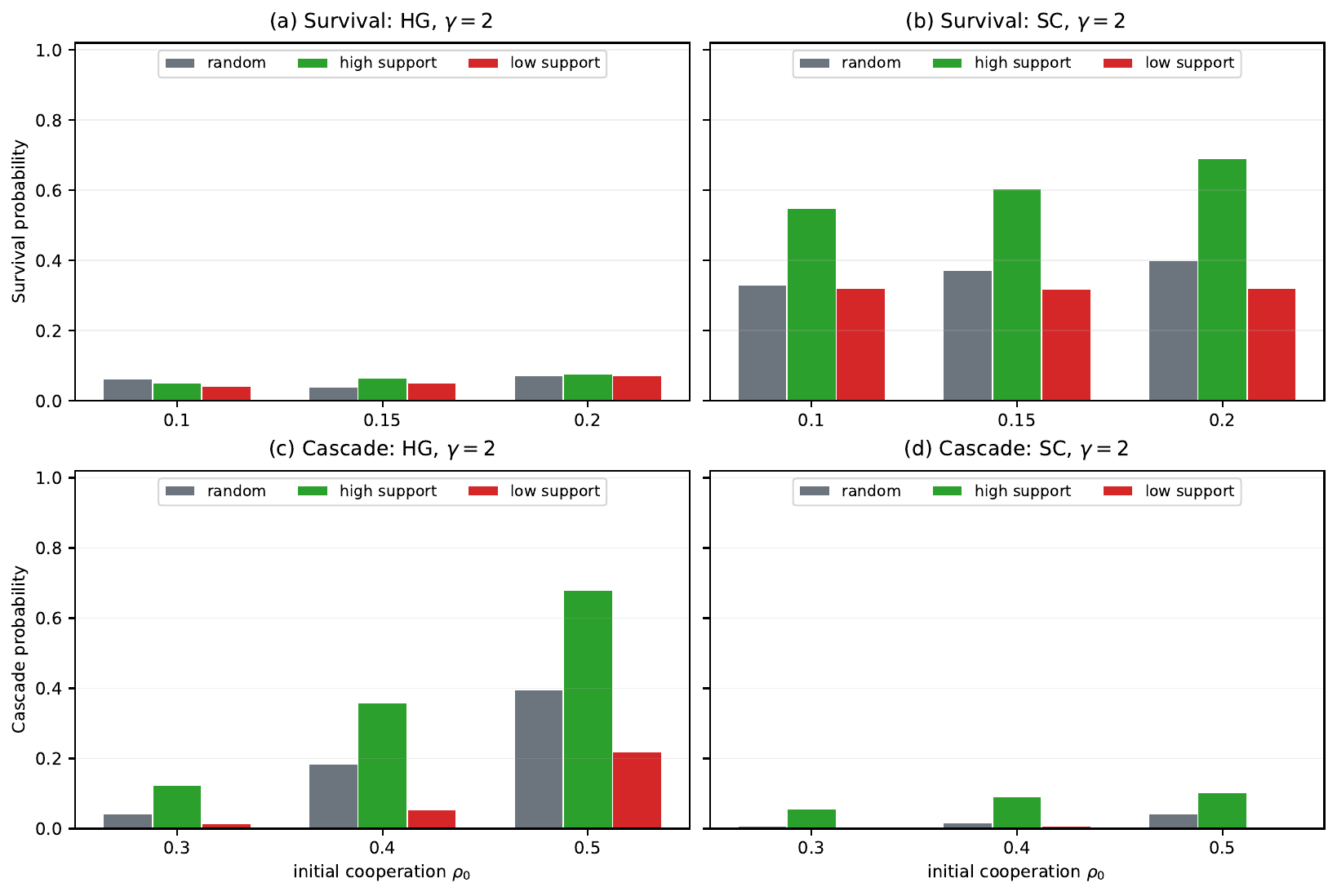}
\caption{Survival and cascade probabilities of cooperative seeds as a function of initial cooperation density $\rho_0$ in the convex regime
$\gamma=2$, shown for the HG [(a), (c)] and the SC [(b), (d)].
Three seed-placement protocols are distinguished by color: random (gray),
high-support (green), and low-support (red). (a, b)~Survival measured at
$\rho_0=0.1$, $0.15$, $0.2$. (c, d)~Cascade measured at
$\rho_0=0.3$, $0.4$, $0.5$.} \label{fig:fig6}
\end{figure}

For a given initial cooperation density $\rho_0$, we set the number of
initial cooperators to $N_C=\rho_0 N$ and compare three seed-placement
protocols: (i) the random protocol, where the $N_C$ cooperative seeds are chosen
uniformly at random; (ii) the high-support protocol, where the seeds are placed on
the $N_C$ nodes with the largest three-body degree $k_i^{(3)}$; (iii)
the low-support protocol, in which the seeds are placed on the $N_C$ nodes with the
smallest $k_i^{(3)}$.
This choice allows us to test whether cooperative nuclei
located in three-body-rich regions have different fates in HG and SC. After preparing the initial conditions, we evolve the system for $10^5$ update
steps and record the final cooperation level. For each value of $\rho_0$ and
each seed-placement protocol, the results are averaged over $1000$ independent
realizations. A realization is counted as surviving if the final cooperation
level $\rho_C^{\mathrm{final}}$ exceeds the ``survival threshold'', and as undergoing a global cascade if $\rho_C^{\mathrm{final}}$ exceeds the ``cascade threshold''.
We then estimate the survival and cascade probabilities under the three
seed-placement protocols in the regime $\gamma=2$.
We have tested survival thresholds in the range $[0.05,0.20]$ and cascade thresholds in the range $[0.80,0.95]$, and the qualitative trends remain unchanged within these ranges.
As a representative choice, Fig.~\ref{fig:fig6} presents the results
when the survival threshold is $0.15$ and the cascade threshold $0.85$.

Fig.~\ref{fig:fig6}(a),(b) show the survival probability of cooperative
seeds in the HG and SC, respectively. Since the purpose of this measurement
is to test whether small cooperative nuclei can persist locally, we
focus on low initial cooperation densities, namely $\rho_0=0.1,0.15,0.2$. In
this range, survival in the HG is strongly suppressed for all seed-placement
protocols, indicating that sparse cooperative nuclei are generally unstable
in the randomized hypergraph. In the SC, by contrast, the survival
probability is substantially higher and shows a clear dependence on local
triangular support: high-support placement gives the largest survival
probability, whereas low-support placement remains much less effective. This
supports that simplicial closure selectively stabilizes cooperative nuclei
embedded in locally supported triangular regions.

On the other hand, Fig.~\ref{fig:fig6}(c),(d) show the cascade probability
in the HG and SC, respectively. Since this measurement tests whether seeded
cooperation can be amplified into a near-system-wide cooperative state, the
initial cooperative set must be finite rather than too sparse. We therefore
use larger initial cooperation densities, $\rho_0=0.3,0.4,$ and $0.5$. As we
can see, in the HG, the cascade probability increases rapidly with $\rho_0$,
especially for high-support placement. This behavior is consistent with a
nucleation-like picture of the first-order transition, in which cooperation
remains unstable below a critical seed size but can be amplified into a
global cooperative outcome once a sufficiently large nucleus is formed. In
the SC, however, cascade events remain rare, even for high-support
placement. This means that, although simplicial closure does not
completely eliminate global cascades, it strongly reduces the probability
that locally persistent cooperative nuclei are converted into system-wide
cooperation.

Having verified the dual role played by simplicial closure, we can now
understand why it has a qualitative effect only in the explosive-transition
regime. In the explosive regime of the randomized HG, cooperation proceeds
via a global nucleation-and-cascade process: once a cooperative nucleus
exceeds a critical size, effectively delocalized three-body feedback drives
a system-wide cooperative cascade. Simplicial closure disrupts this
all-or-none mechanism by localizing the feedback: it stabilizes small
cooperative nuclei embedded in closed triangles---thereby enhancing their
local survival---while simultaneously confining their propagation to
overlapping triangular neighborhoods---thereby suppressing global cascades.
The net outcome is the fragmentation of the HG's compact two-basin structure
into a broad ensemble of spatially distributed metastable states. By
contrast, in the continuous-like regime, no global nucleation threshold
exists; cooperation grows incrementally through spatially intermixed local
configurations. Consequently, the same dual role acts only as a quantitative
perturbation, shifting or broadening the transition without restructuring
the underlying attractor landscape.

\section{Breaking simplicial closure}

The previous sections showed that simplicial closure qualitatively affects
cooperation dynamics only in the explosive-transition regime, where it
fragments the global bistability of the randomized HG into a broad set of
locally persistent metastable outcomes. Having identified the corresponding
mechanism, we now examine whether this fragmentation is robust to partial
breaking of simplicial closure. To this end, we introduce a closure-rewiring
interpolation between SC and HG. Starting from the original SC, each
three-body interaction is randomly rewired with probability $p$, while
the total number of three-body groups is kept fixed. Thus, $p=0$ corresponds
to the original SC, whereas increasing $p$ progressively destroys the local
triangular constraint and makes the higher-order interactions more HG-like.

We focus on the convex payoff regime $\gamma=2$. For clarity, we present
only two representative diagnostics: the final-state overlap distribution $%
P(Q)$, which probes the structure of the attractor landscape, and the time
evolution of the largest cooperative component $S_{\max}^{CC}(t)/N$, which
probes whether cooperative domains remain localized or grow into system-wide
cascades. Other diagnostics, including the final cooperation distribution
and hysteresis scans, show the same crossover trend and are therefore not
shown here.

\begin{figure}[htbp]
\centering
\includegraphics[width=0.8\textwidth]{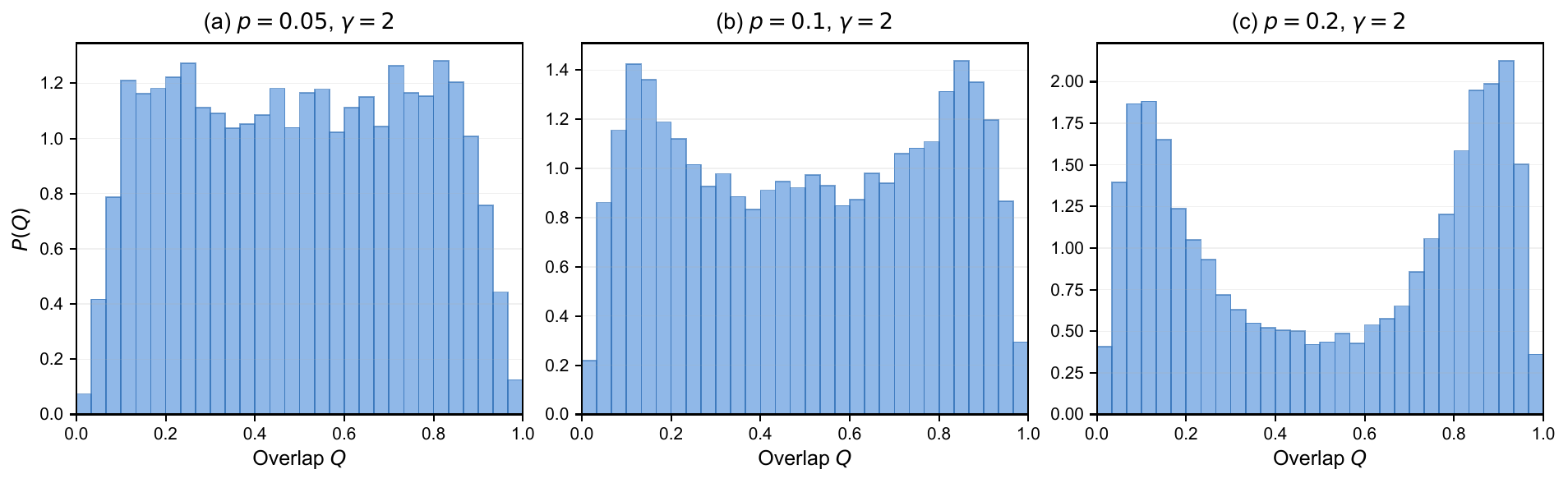}
\caption{Final-state overlap distributions $P(Q)$ in the closure-breaking interpolation for the convex regime $\gamma=2$.  The parameter
    $p$ denotes the probability of breaking simplicial closure by rewiring
    three-body interactions while keeping their total number fixed.  For
    $p=0.05$, the distribution remains close to the SC case. For
    $p=0.1$, the distribution starts to accumulate near both low and high
    overlaps, and when $p$ grows to $0.2$, a clearer bimodal
    structure emerges, showing the recovery of an HG-like two-basin organization.}
\label{fig:fig7}
\end{figure}

Fig.~\ref{fig:fig7} shows how the final-state overlap distribution $P(Q)$
changes as simplicial closure is progressively broken. For $p=0.05$, $P(Q)$
remains broadly spread over the whole range of overlap values, close to the
SC regime. At $p=0.1$, the distribution starts to deviate from this broad
profile and accumulates more weight near low and high overlaps, signaling a
crossover toward HG-like behavior. When $p$ is further increased to $0.2$, a
clearer bimodal structure emerges, indicating that partial breaking of
simplicial closure already restores much of the compact two-basin
organization characteristic of the randomized HG.

\begin{figure}[htbp]
\centering
\includegraphics[width=0.8\textwidth]{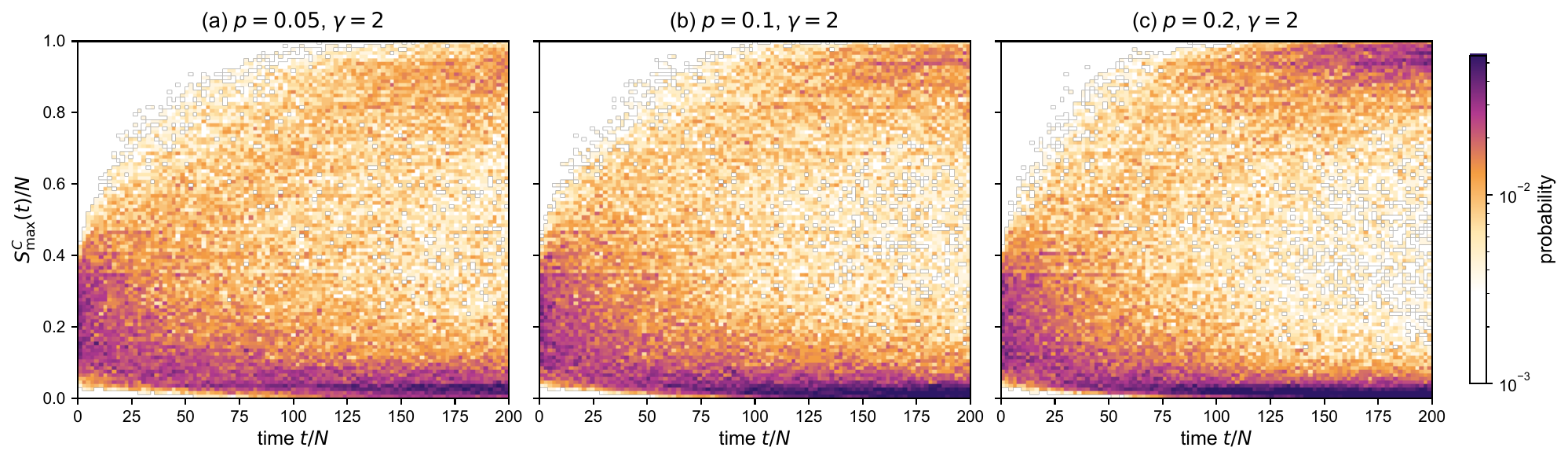}
\caption{Time evolution of the probability distribution of the normalized largest
cooperative component $S_{\max}^{CC}(t)/N$ under partial breaking of
simplicial closure for $\gamma=2$. At a small closure-breaking probability,
$p=0.05$, cooperative components remain broadly distributed over intermediate
sizes, reflecting SC-like localized persistence. As $p$ increases, the
distribution becomes more polarized between low-cooperation trajectories and
system-spanning cooperative trajectories, indicating a gradual recovery of
HG-like cascade dynamics.} \label{fig:fig8}
\end{figure}

The crossover is also visible in the dynamics of the largest cooperative
component. Fig.~\ref{fig:fig8} shows the time-resolved distribution of $%
S_{\max}^{CC}(t)/N$ for the same values of $p$. For $p=0.05$, the
distribution remains broadly populated over intermediate component sizes,
indicating that cooperative domains are still largely localized, as in the
SC regime. At $p=0.1$, the probability gradually shifts toward larger
components while intermediate sizes remain visible, corresponding to a
crossover between localized persistence and cascade-like growth. For $p=0.2$%
, the distribution becomes more polarized and the middle part become less
pronounced, indicating partial breaking of simplicial closure progressively
restores the HG-like dynamics, consistent with Fig.~\ref{fig:fig7}.

Taken together, Figs.~\ref{fig:fig7} and \ref{fig:fig8} show that even
partial breaking of simplicial closure gradually restores the HG-like
two-basin organization and cascade dynamics. This also confirm that the
fragmentation of bistability observed in the SC is directly induced by
simplicial closure.

\section{PGG with Triple-product payoff}

The analysis above has focused on the normalized scale-effect payoff.
However, the mechanism identified here is not tied to this particular
functional form. What matters is whether the payoff rule can generate an
explosive transition in the randomized HG, because only in this case can
simplicial closure qualitatively alter the transition by disrupting the
conversion of cooperative nuclei into system-wide cascades. To verify this
point, we consider an alternative three-body payoff that selects fully
cooperative triads more strictly. Specifically, we employ the
triple-product-form payoff
\begin{equation}
u_i^{(3)}(i,j,k) = 3\alpha r (s_i s_j s_k) - s_i ,  \label{eq:product_payoff}
\end{equation}
where $s_i\in\{0,1\}$. The product $s_i s_j s_k$ is nonzero only when all
three players cooperate, so the three-body benefit is generated exclusively
by fully cooperative triads. Economically, this corresponds to a
complementary group task in which every participant is indispensable.
Dynamically, it provides an extreme form of higher-order positive feedback:
mixed triads receive no three-body benefit, whereas fully cooperative triads
are strongly rewarded. It is therefore expected to produce a
first-order-like transition in the randomized HG and provides a suitable
test of whether the fragmentation effect of simplicial closure extends
beyond the power-law scale-effect payoff.

We test this product-form payoff on the same HG and SC representations used
above, keeping the network construction and dynamical parameters unchanged.
For clarity, we only present three representative diagnostics in Fig.~\ref%
{fig:fig9}: the average cooperation curve, the final-state distribution, and
the final-state overlap distribution. These diagnostics are sufficient to
determine whether the product payoff produces an explosive transition in the
randomized HG and whether simplicial closure fragments the corresponding
bistability.

\begin{figure}[htbp]
\centering
\includegraphics[width=0.85\textwidth]{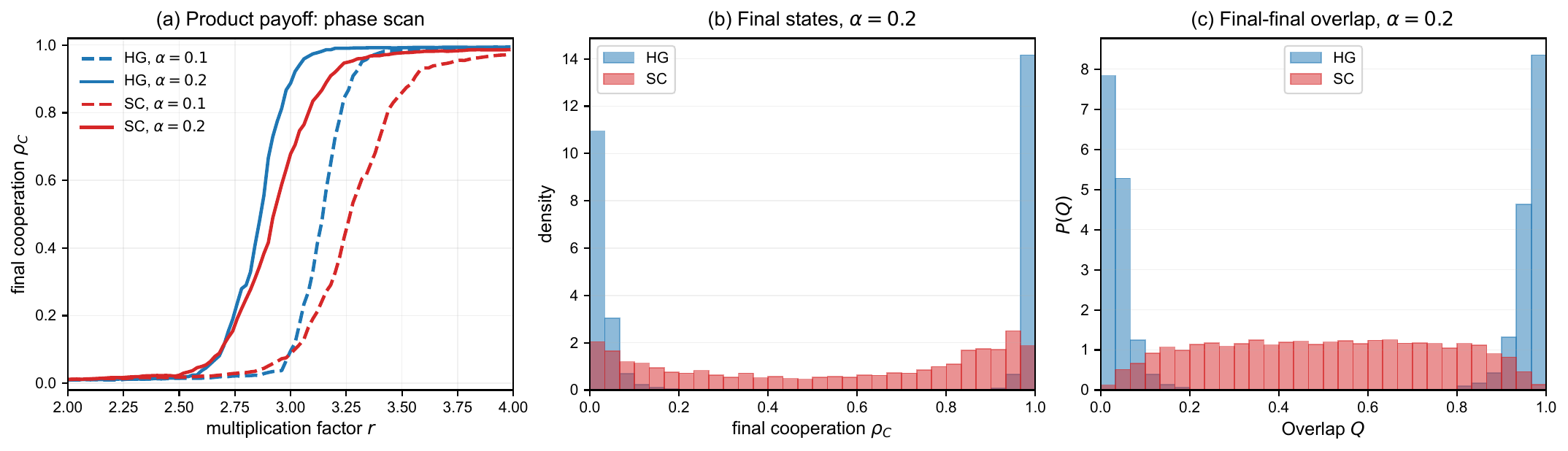}
\caption{Cooperation dynamics under the strict product-form three-body payoff.
(a) Final cooperation level $\rho_C$ as a function of the multiplication
factor $r$ for HG and SC with $\alpha=0.1$ and $\alpha=0.2$.
(b) Distribution of final cooperation levels for $\alpha=0.2$.
(c) Distribution of final-state overlap $Q$ for $\alpha=0.2$.
The randomized HG shows sharp transitions, low/high bistability, and a
two-basin-like overlap structure, whereas the SC produces smoother transitions,
intermediate final cooperation levels, and broadly distributed overlaps. These
behaviors are fully consistent with those observed previously in the convex
$\gamma=2$ regime.} \label{fig:fig9}
\end{figure}

As shown in Fig.~\ref{fig:fig9}(a), the triple-product payoff generates
sharp cooperation transitions in the randomized HG, consistent with its
strong selection for fully cooperative triads. Increasing $\alpha$ shifts
the transition to smaller values of $r$, as expected. The SC transitions are
generally smoother and broader, reminiscent of the convex regime of the
power-law payoff. The final-state distributions at the transition points ($r_c=2.86/2.94$
for $\alpha=0.2$ on HG/SC) in Fig.~\ref{fig:fig9}(b)
also show the expected structures: in the HG, realizations
concentrate near the low- and high-cooperation states forming the bimodal structure,
whereas in the SC the distribution spreads over a wide range of intermediate cooperation levels,
similar to Fig.~\ref{fig:fig2}(b). The corresponding overlap distribution in Fig.~\ref%
{fig:fig9}(c) further confirms this difference at the configuration level:
the HG displays a two-basin-like overlap structure, while the SC exhibits a
broad overlap distribution indicating many distinct metastable final
configurations.

Thus, we confirm that the qualitative effect of simplicial closure is not
tied to a specific payoff form, but is instead a generic feature for
explosive cooperation transitions.

\section{Conclusion and Discussion}

In this work, we investigated how the representation of higher-order
interactions affects evolutionary game dynamics. By studying controlled
higher-order PGGs with multiple types of payoff on HG and
SC, we show that the influence of simplicial closure is
selective. Specifically, when the corresponding randomized HG exhibits a
continuous-like cooperation transition, imposing simplicial closure mainly
produces quantitative changes, without changing its continuous-like nature.
In sharp contrast, when the randomized HG exhibits an explosive
first-order-like transition, simplicial closure fragments the compact
low/high bistability into a broad ensemble of metastable final states. In
other words, HG and SC representations become \emph{qualitatively}
inequivalent when the payoff rule supports cooperative nucleation and
cascade-like expansion.

Detailed microscopic diagnostics and seed experiments further reveal why
this inequivalence emerges only in the explosive-transition regime. In
randomized HG, sparse cooperative nuclei are generally fragile, but once a
sufficiently large nucleus is formed, effectively delocalized three-player
feedback can amplify it into a system-wide cooperative cascade. This
produces the compact two-basin structure characteristic of the
first-order-like transition. In SC, by contrast, simplicial closure plays a
dual role: it enhances the local survival of cooperative nuclei embedded in
closed triangular neighborhoods, but also confines their propagation through
the same local triangular organization. Consequently, the global bistability
observed in HG is not simply suppressed, but fragmented into a broad
ensemble of locally persistent metastable states. In continuous-like
transition regimes, by contrast, cooperation spreads incrementally rather
than through nucleation and cascade conversion; therefore the same
closure-induced localization mainly acts as a quantitative structural
perturbation instead of reorganizing the attractor landscape.

It is worth noting that the HG--SC distinction identified here is different
from those reported in synchronization and
contagion dynamics. In synchronization dynamics\cite{zhang2023}, the
difference between HGs and SCs is mainly expressed through higher-order
degree heterogeneity and cross-order degree correlations, which reshape the
spectral stability of the synchronized state. In higher-order contagion\cite%
{maia2026}, the key distinction is instead associated with nesting or
embedding, namely the extent to which lower-order interactions are contained
within higher-order ones, thereby controlling activation thresholds and the
emergence or suppression of hysteresis. In the present evolutionary game
setting, however, the relevant distinction is the spatial fate of
cooperative feedback: whether locally formed cooperative nuclei are
amplified into system-wide cascades or remain confined as metastable local
domains. This comparison suggests that the consequence of a higher-order
representation cannot be inferred from simplicial closure alone; rather, the
meaning of the HG--SC difference is dynamical-process dependent. In this
sense, the present work contributes to a broader understanding of how
higher-order representations shape collective dynamics in complex systems.

Several directions follow naturally from this study. First, it would be
useful to develop reduced theoretical descriptions that relate the
fragmentation of bistability to measurable structural quantities, such as
triangle overlap or the distribution of connected
cooperative domains. Second, the present analysis can be extended to other
forms of higher-order evolutionary games, including threshold public goods
games, or games with heterogeneous costs or nonlinear
benefit-sharing rules. Such extensions would provide a broader setting for
applying the mechanism identified here to different cooperation scenarios.

\end{document}